# Fabrication of Metal Nanoscale Devices on Insulating Membranes by High-Resolution Atom Ablation

M. D. Fischbein & M. Drndić

*Department of Physics and Astronomy, University of Pennsylvania, Philadelphia, PA 19104*

## Abstract

Transmission electron beams (TEBs) have long been used to study and manipulate materials at nanometer scales. In the late 1970's, Cherns demonstrated surface pitting in Au films upon exposure to a 1MeV TEB [1,2] and observed crystal dislocations in quartz [3]. Soon after, electron beam irradiation was used to drill nanoholes and lines in NaCl crystals [4], alumina sheets [5,6,7], $CaF_2$ and MgO [7]. More recent examples include the drilling of nanoholes in silicon [8], stainless steel [9], and in $Si_3N_4$ and $SiO_2$ membranes [10,11,12]. In this Letter we demonstrate a new and highly flexible application of TEB-based fabrication to produce intricate metal geometries and fully integrated devices, with sub-10 nm features, on silicon nitride membranes. Arbitrary metal patterns may be "carved out" with sub-nanometer accuracy by ablating evaporated (Al, Ni, Cr, Ag, Au) metal films with the ~ 5 Å diameter beam of a high resolution transmission electron microscope. *In situ* imaging of the ablation action allows for real-time feedback control. Specific examples presented here include nanorings, nanowires with tailored curvatures and multi-terminal devices with nanoislands or nanoholes between the terminals. Importantly, these nanostructures are fabricated at precise locations on a chip and seamlessly integrated into large-scale circuitry. The combination of high resolution, geometrical control and yield make this fabrication method rather unique and highly attractive for many applications including nanoelectronics and molecular translocation.

Physical, chemical and biological properties of diverse systems depend on electron motion, fluid motion and/or chemical reactions that occur at nanometer scales. For instance, electrons typically travel a few nanometers at room temperature before scattering inelastically in metals [13] or flipping their spin in ferromagnetic metals [14] and their transition into the superconducting phase becomes sensitive to size at ~ 10 nm [15,16,17,18]. Nanometer scale fluid flow deviates significantly from bulk flow [19] and, in fact, water confined in nanometer scale volumes crystallizes at room temperature [20]. Control over molecular motion in sub-10 nm wide channels is important for studies of molecular-scale mass transport and for nanofluidic applications in genomics [21]. In pursuing nanoscale science many different and original approaches have been undertaken to fabricate small structures. However, top-down fabrication of sub-10 nm scale devices with high reproducibility and yield is still challenging.

The use of transmission electron beams (TEBs) for nanofabrication was first demonstrated in the late 70's and early 80's. Cherns was the first to demonstrate that a high-energy TEB can induce "transmission sputtering" [1] in Au, and proposed a many-body collision model to explain this effect by surface diffusion and agglomeration of surface vacancies[1,2]. This pioneering work was followed by other researchers who were able to drill holes and lines in a variety of crystalline and amorphous solids[3-7].

Upon electron irradiation of the sample, high-energy electrons may lose a portion of their kinetic energy via inelastic scattering processes in the solid, resulting in various effects including sputtering, amorphization, sublimation and desorption. TEB-induced atomic rearrangement has been observed Au piezoelectric point contacts[22], Au nanoparticles[23] and freestanding Au nanobridges suspended from Au (001) films.[24] TEBs have been used to observe the sublimation of nanometer Ag particles at 950 K [25], to melt Al-Si particles [26], to transform diamond nanoparticles into onion-like carbon [27] and to reduce $SiO_2$ to Si [28]. In addition to drilling nanoholes, as mentioned above[1-12], TEBs have recently attracted renewed interest for nanofabrication. They have been used to break Au and Pt wires [29], to join single-wall carbon nanotubes[30] and to "weld" and inscribe synthesized silicon and metal nanowires [31] on holey carbon grids.

Here we report a new TEB-based method to controllably ablate evaporated metal films, pre-patterned with electron beam lithography (EBL) on silicon nitride ($Si_3N_4$) membrane substrates, to produce fully integrated device components with near atomic precision that can be easily connected to larger circuitry. We show that metal nanostructures can be carved out with sub-nm precision while preserving the underlying insulating platform on which devices are fabricated (see Methods). To convey the unparalleled flexibility of this resist-free TEB-based "ablation lithography" (TEBAL), we have chosen several examples of TEBAL fabricated structures that embody this versatility and likewise, to our knowledge, have not been demonstrated at this size scale with any other technique. Figure 1 shows three example structures made from a Ni/Cr alloy: a nanoring (Fig. 1a) with inner and outer diameters of 6 and 30 nm, respectively, a three-terminal single-electron transistor (Fig. 1b) and a serpentine wire (Fig. 1c) having a width of 7 nm and radius of curvature of ~1 nm at the three bends. A corresponding schematic is also shown for each structure for clarity. Each structure was intentionally left connected to its "parent leads" to demonstrate the ease of their integration into complete circuits. The feature sizes were measured with the same Gatan Digital Micrograph Image Software that was used to obtain the HRTEM images. The images were not altered or processed in any way after being captured by the CCD camera of the HRTEM (i.e., they are "raw"). Specifically, the contrast, brightness and sharpness in the images are exactly as they appear to the user during the fabrication. Post-fabrication, the structures were imaged several times over a period of weeks and displayed no sign of changing or "relaxing" into different shapes over time. We note that TEBAL fabrication has the added advantage of maintaining a resistance-free contact between a nanostructure and its leads, in contrast to bottom-up nanostructures which need to be first located on a chip and then contacted to larger circuitry, all after their fabrication.

Figure 2 shows an image gallery of metal structures made with TEBAL on $Si_3N_4$ membranes. A nanowire is shown at several stages of its formation (Fig. 2a-c) and several additional examples of finished nanowires are also shown (Fig. 2d-f). Two-terminal electrical characterization of the 3.5 nm wide nanowire shown in Fig. 2e under vacuum (pressure < $10^{-6}$ Torr) from room temperature down to ~ 4 K displayed Ohmic resistance of ~50 kΩ with an applied bias of 50 mV (Fig. S1). Also demonstrated are nanochannels (Fig. 2 g-k), nanoparticles (Fig. 2l), nanoholes (Fig. 2m) and nanogaps (Fig. 2n-o). Nanowires, nanogaps and nanochannels, whose aspect ratios and geometries can be finely tuned, are particularly interesting for studies of superconductivity, molecular electronics, nanofluidics and nanoparticle manipulation with electromagnetic fields.

Figure 3 shows another type of device made with TEBAL which combines nanogaps and nanoholes. It consists of a pair of electrodes roughly 2 nm apart and a nanohole drilled into the membrane exactly between them. The third electrode enables this device to be used as a nanogap field effect transistor (NGFET) for a range of molecular electronic applications. The source-drain leakage current, for the device in Fig. 3c was ~ 20 fA at 100 mV (Fig. S2). The current sensitivity is therefore very high allowing the study of even highly insulating molecules. As with all TEBAL fabricated structures, the geometrical features are easily reproduced as many times as desired. To illustrate the reproducibility, we show three of these devices (Fig. 3a-c), which were all fabricated on the same 50 x 50 μm$^2$ membrane window (Fig. 3e). Several groups have demonstrated the fabrication of nanoholes with controlled sizes in insulating membranes by using TEBs[10,11,12] and focused ion beams (FIB)[32]. Nanoholes were useful for manufacturing single-molecule detectors[33-39] and were used to study translocation of DNA molecules[11,35,37] and carbon nanotubes[40] by measuring ionic current between macroscopic electrodes as molecules passed through the nanohole. Solid-state nanoholes were recently used to study the folding effects in double-stranded DNA[35,37]. These and other advancements (see Supplementary References) made it possible to envision future applications of nanoholes, including DNA size determination and sequencing. The devices we show in Figure 3 are comprised of additional electrodes and gates, positioned precisely near the hole. This TEM-based advancement in fabrication adds versatility to nanohole devices for single molecule detection and could provide additional information about the translocation dynamics and the conformational states of DNA molecules passing through the nanoholes. Specifically, the source and drain electrodes could both act as STM tips, providing a near-field source and probe that are aligned by construction. By applying a potential difference across the electrodes, electrical current (or capacitance) could be monitored to characterize the molecule transversely during its translocation through the nanohole (illustrated in Fig. 3f). Lagerqvist *et al.*[41] have recently performed molecular dynamics and quantum-mechanical current calculations for this situation and calculate the difference in electrical current signals of different base pairs to be rather large, of the order of 0.1 nA. However promising, there are several nontrivial experimental details to be surmounted and the feasibility of this approach will be ultimately determined by future experiments.

In conclusion, we have demonstrated the precision, versatility and reproducibility of controlled atom ablation via the imaging beam of a HRTEM for the manufacturing of ultra-small metal device features on insulating membranes. The applications of TEBAL are far reaching and include the areas of nanoelectronics, superconductivity, molecular translocation, nanofluidics and nanoparticle manipulation with electromagnetic fields.

**METHODS**

The TEBAL process is outlined in Figure 4. In order to perform TEBAL we first introduce the initial metal that will be nano-sculpted by the ablating beam. This initial metal is prepared on a thin membrane that is itself essentially transparent to the electron beam. The parallel fabrication of many such membranes is well documented in existing literature [42]. We use membranes made of low-stress amorphous $Si_3N_4$. $Si_3N_4$ membranes are an excellent platform for device manufacturing; they have a dielectric constant of ~ 6.5 – 7.2 and a breakdown voltage of ~ $10^7$ V/cm, making them an attractive alternative dielectric material compared to $SiO_2$ [43]. By using existing methods such as electron beam and photo lithography, we pattern thin strips of metal onto one side of the membrane surface. The resolution achieved on $Si_3N_4$ membranes with electron beam lithography (EBL) is enhanced significantly due to the absence of electron backscattering[44] and nanogaps can be made directly with EBL [45]. Metal films were evaporated using standard thermal evaporator procedures at a rate 0.3 nm/s and pressure < $10^{-6}$ torr. We are therefore able to start with metal features that are already small compared to what is usually achieved with EBL on $SiO_2/Si^{++}$ substrates and then proceed with TEBAL to make structures inaccessible with EBL only (e.g., structures in Fig. 1). The membrane device is then loaded onto a TEM device holder (Fig. 4a) and entered into the low-pressure (< $10^{-8}$ Torr) chamber of a HRTEM at room temperature. Upon exposure to the intense TEB, metal is knocked-off from the substrate, leaving the desired pattern.

Using the standard imaging mode of the microscope, the region of the initial metal to undergo TEBAL is located. Next, the magnification is increased and focused. During TEBAL the user is actually able to see the ablating effect of the caustic spot in real time because of the low-intensity illumination of the beam surrounding this high-intensity center region. The time between exposing the metal to the condensed beam and the ablation is on the order of seconds, though the exact time required will depend on the material being ablated and the microscope conditions. For all of the metals that we have explored (Ag, Ni, Cr, Al & Au), the ablation effect seems to stop almost immediately after the current density around the ablated area is reduced.

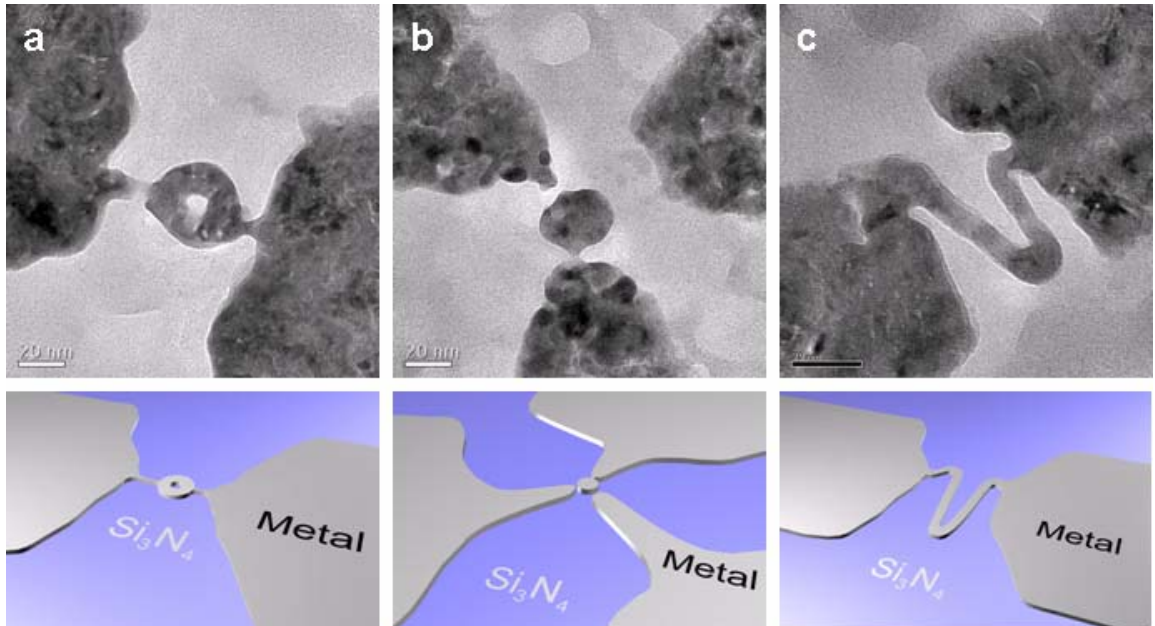

**Figure 1** Example structures to demonstrate the flexibility of TEBAL. Each of the three structures shown in the TEM images is accompanied by a schematic (below) showing the fabrication by TEBAL. **a**, Nanoring with outer radius of ~15 nm and inner radius of ~3 nm (scale = 20 nm). **b**, Three-terminal electronic device: source and drain leads are coupled to a ~ 25 nm diameter metallic island and a gate electrode is positioned nearby (scale = 20 nm). The rate limiting tunneling barrier (upper junction) is a 2 nm gap. **c**, Serpentine wire with 7 nm width (scale = 20 nm). All lengths were measured with Gatan's Digital Micrograph Image Analysis software.

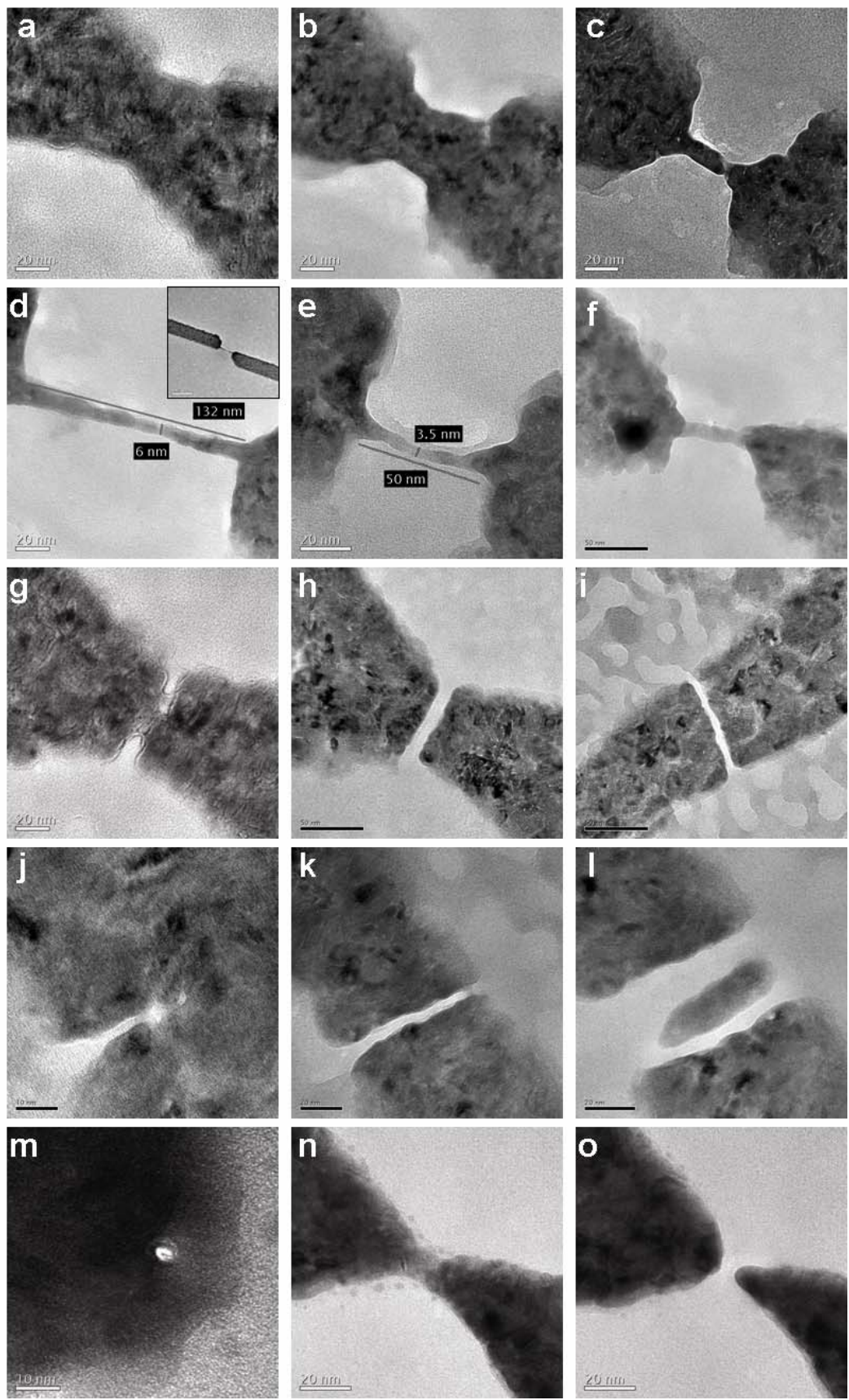

**Figure 2** TEBAL example gallery.  **a-c**, TEBAL fabrication of a nanowire made of a Ni/Cr alloy (all scale bars = 20 nm). **d-f**, Additional examples of Ni/Cr alloy nanowires. The width and length (W,L) of the nanowires in nanometers are (6, 132), (3.5, 50) & (7, 50), respectively (scale bars = 20 nm, 20 nm & 50 nm).  **g**, "Bottleneck" structure (scale bar = 20 nm) **h**, Nanochannel ~ 7 nm wide. (scale bar = 50 nm) **i**, Nanochannel 3 nm wide, 70 nm long. (scale bar = 50 nm). **j-l**, Three stages of "carving out" a Ni nanoparticle (all scale bars = 20 nm).  **m**, 3 nm diameter nanohole drilled into Cr. (scale bar = 10 nm).  **n&o**, Au wire with narrow constriction before and after removing debris and making a clean 5 nm nanogap with TEBAL (scale bars = 20 nm).

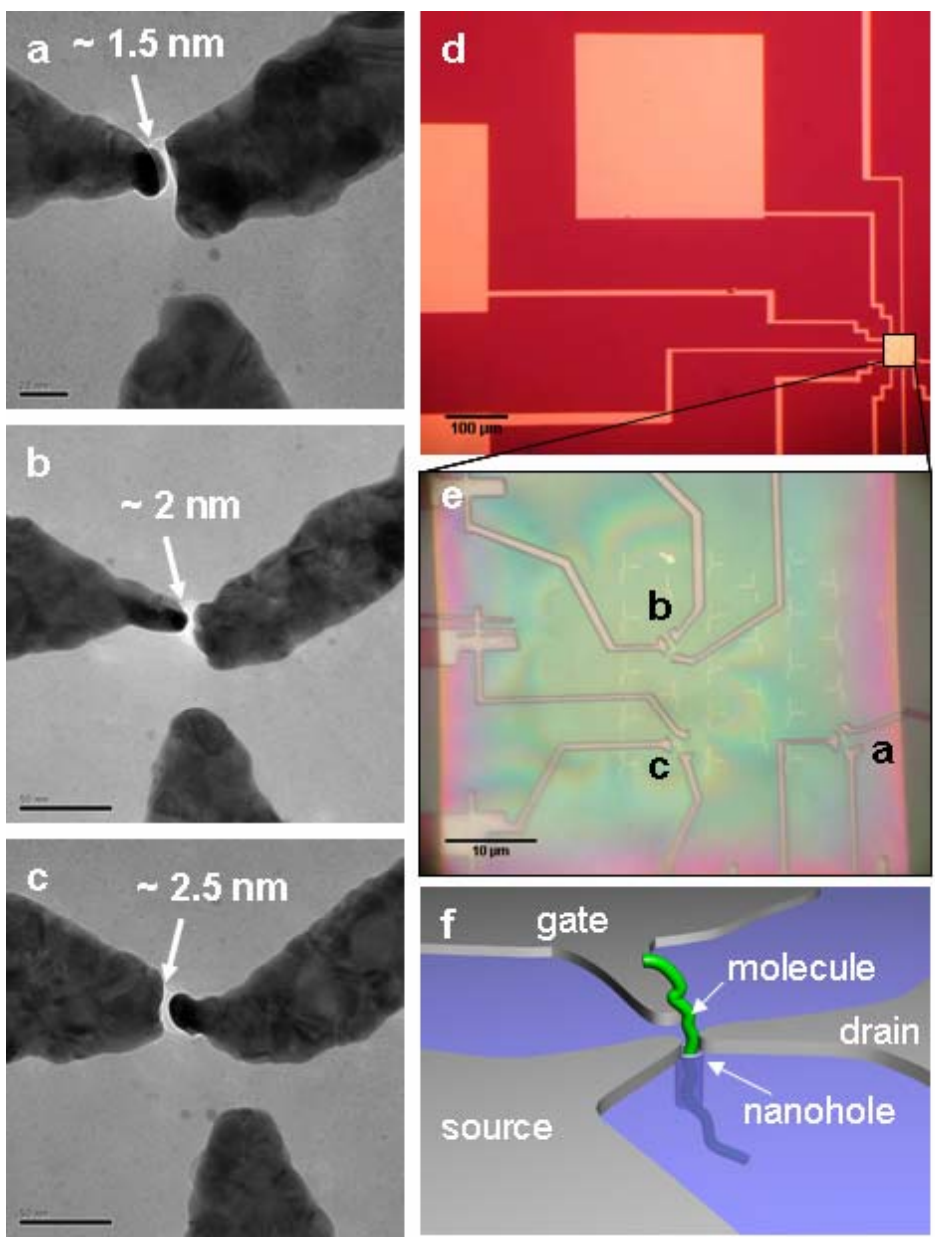

**Figure 3** Devices for molecular detection and analysis. **a-c**, Source, drain and gate devices with nanohole for molecular translocation. (scale bars = 20 nm, 50 nm & 50 nm) **d**, Optical micrograph of the full device containing the three molecular "analyzers". The membrane window is seen in the lower right corner and has wires running from it out to large contact pads. (scale bar = 100 µm) **e**, Optical micrograph of the membrane window showing the three analyzers (a-c) connected to leads. (scale bar = 10 µm) **f**, Schematic illustrating vertical translocation of a molecule through the nanohole while being probed with the source, drain and gate electrodes.

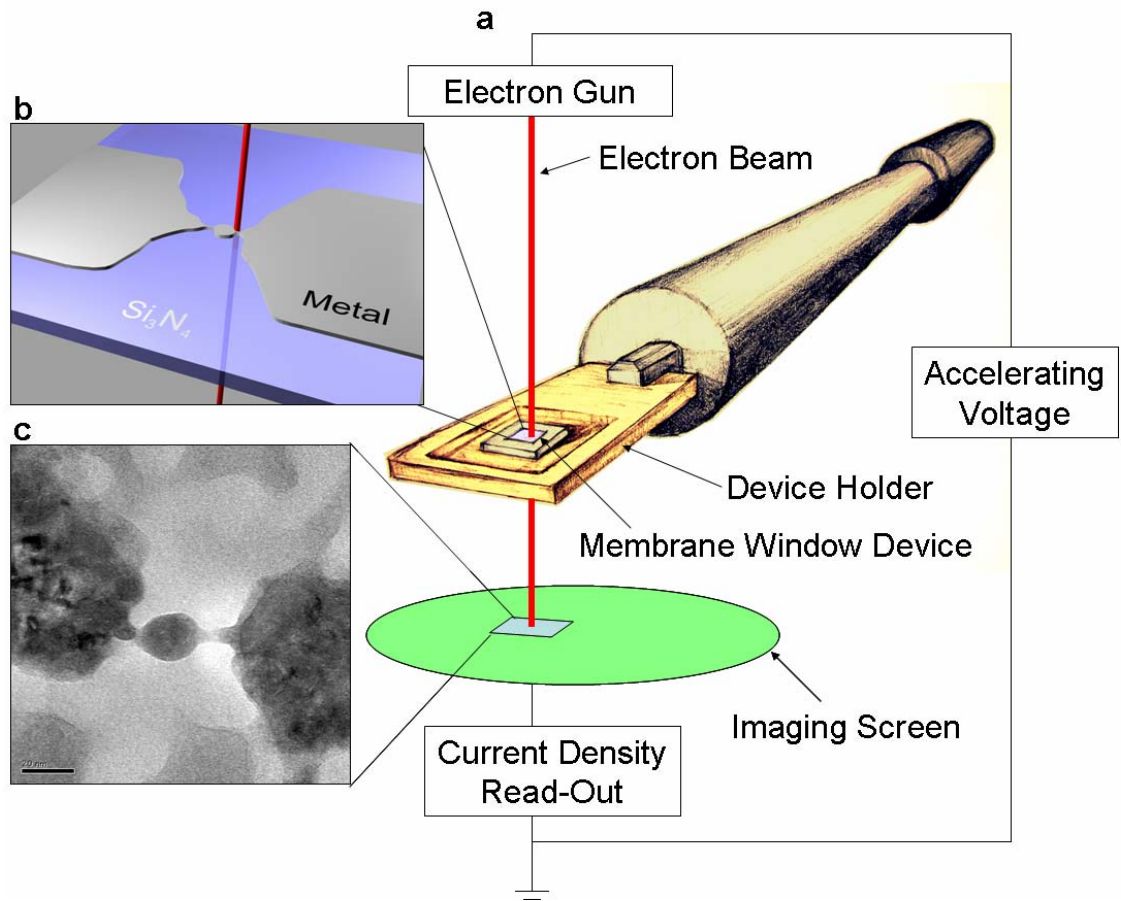

**Figure 4** Apparatus for TEBAL. **a**, A membrane window device with pre-patterned metal lines on its surface is loaded into the TEM environment. The metallized surface of the membrane faces away from the beam source (i.e., towards the imaging screen). A hole (not shown) in the chip holder allows the beam to pass completely through the membrane and reach the imaging screen. **b**, Schematic of the ablation of a pre-patterned wire to achieve a metallic disk that is connected to the initial metal on both sides. **c**, TEM image of an actual 20 nm diameter metallic disk that was nano-sculpted from Cr/Ni with TEBAL and left connected to the parent-leads with two 4 nm wide nanowires (scale = 20 nm). The imaging screen enables real-time visual feedback for a user or real-time current density feedback for computerized control.

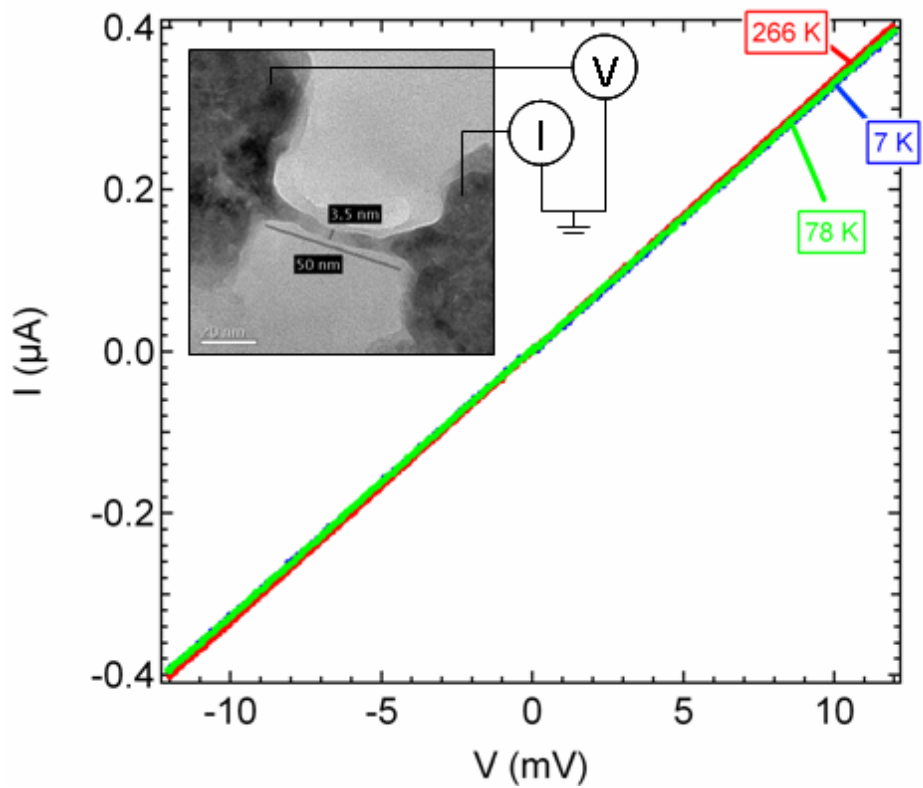

**Figure S1** Current-voltage measurements of 3.5 nm wide nanowire shown in Fig 2e. Data shown for measurement temperatures 266, 78 and 7 K. Inset: TEM image of the nanowire (scale = 20 nm).

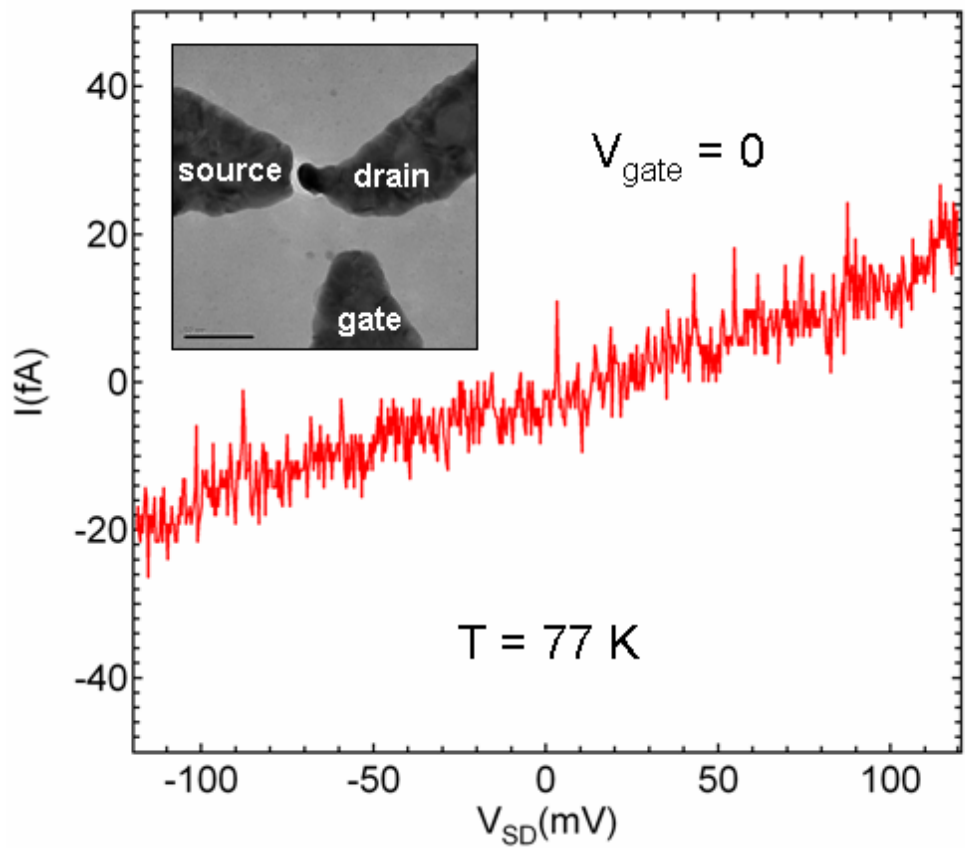

**Figure S2** Background source-drain current-voltage signal of molecular analyzer shown in Fig. 2c. Inset: TEM image of the device (scale = 50 nm).

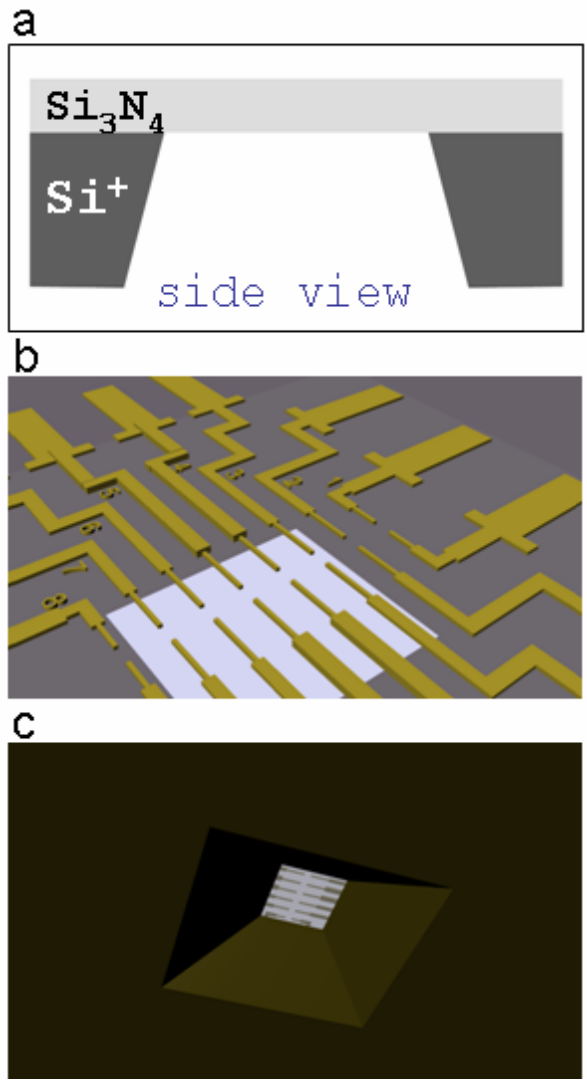

**Figure S3** Schematics of Si$_3$N$_4$ membrane window device. Membranes can be etched down to at least 10 nm thick with reactive ion etching. **a**, Cross-sectional view illustrating the Si$_3$N$_4$ membrane suspended over an open region of Si$^+$. **b**, Top-view (perspective) illustrating electronics on the window (white square) and continuing away from it towards the perimeter of the device. **c**, Bottom-view (perspective) illustrating the Si$_3$N$_4$ membrane window with electronics on its surface (as in Fig. S3b) as seen from the under-side of the chip.

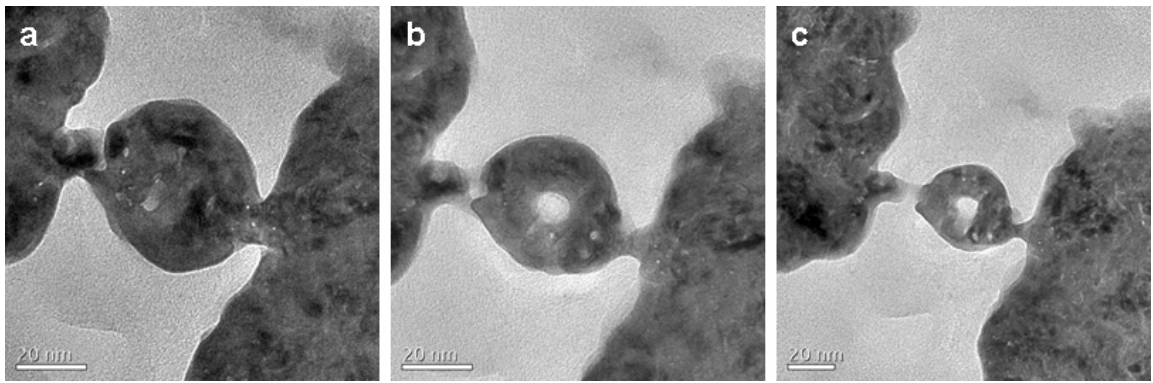

**Figure S4** TEBAL fabrication of the nanoring in Fig. 2c. **a**, Initial wire **b**, TEBAL fabrication of a ~ 45 nm diameter disk. **c**, Disk reduced to ~ 39 nm diameter and a 6 nm diameter hole made in its center. **d**, Final nanoring with ~ 37 nm outer diameter and 6 nm inner diameter. All scale bars = 20 nm.

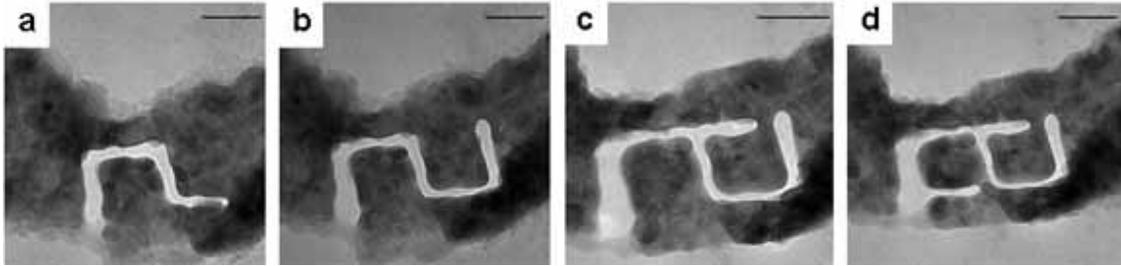

**Figure S5** Several steps towards making a pair of squares in Ag. **a**, First features are composed of four joint line segments, each at right angles to its neighbor. **b**, Addition of line segment at 90 degrees to the right most line segment. **c**, Right square is defined and left connected to the parent material by a short nanowire. **d**, Left square is defined and left connected to the parent material by a short nanowire. All scale bars = 50 nm.

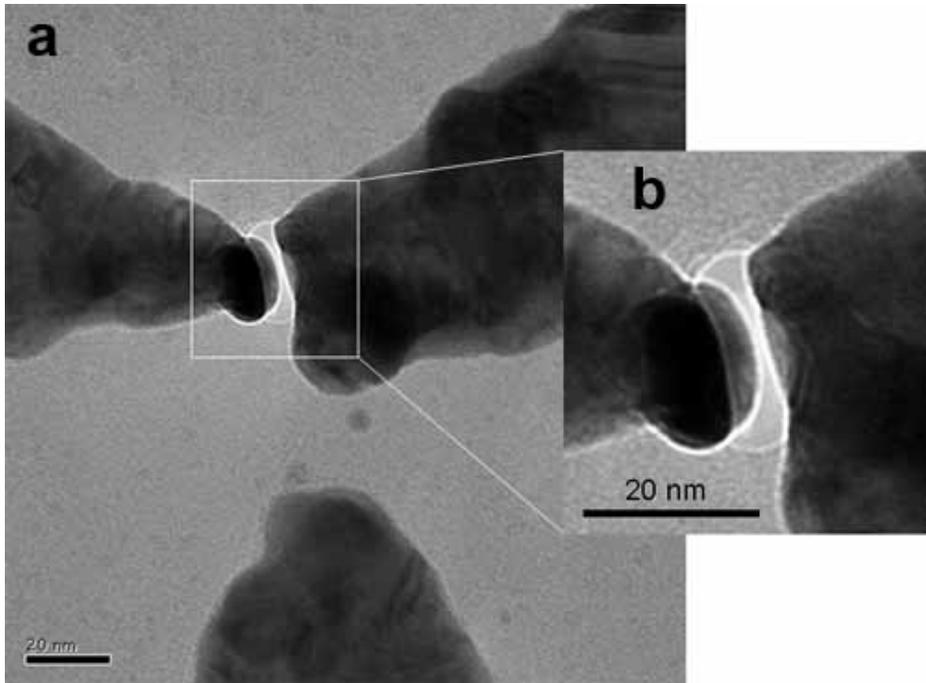

**Figure S6** Larger view of the device in Fig. 3c showing the nanohole in the nanogap. **a**, Figure 3c reproduced at a larger scale. **b**, Close-up view of the nanogap region showing the hole in the nitride membrane between the electrodes.